# Synergy between 6G and AI: Open Future Horizons and Impending Security Risks

*Elias Yaacoub*

*Computer Science and Engineering Department, Qatar University, Doha, Qatar.*

*Emails: eliasy@ieee.org*

*Abstract* – **This paper investigates the synergy between 6G and AI. It argues that they can unlock future horizons, by discussing how they can address future challenges in healthcare, transportation, virtual reality, education, resource management, robotics, in addition to public safety and warfare. However, these great opportunities come also with greater risk. Therefore, the paper provides an overview of the security risks and challenges, along with possible mitigation techniques.**

## 1. INTRODUCTION

With the standardization of 5G becoming a reality, 6G research is increasing at a significant pace (Gatherer, 2018; Saad et al., 2020; RF Wireless World, 2020; Latva-aho and Leppänen, 2019; Giordani and Zorzi, 2019; Giordani and Zorzi, 2020). 5G had identified three main use cases (ITU-R M.2083, 2015):

 - Enhanced mobile broadband (eMBB), which corresponds to increased "human" traffic on the network due to the massive use of multimedia and the increased trend in using gaming and virtual reality (VR) services over the network.

 - Massive machine-type communications (mMTC), which mainly deals with "machine" traffic over the network, thus coping with the billions of sensors and actuators deployed under the internet of things (IoT) paradigm.

 - Ultra-reliability and low-latency communications (URLLC), which deals with highly critical services that are highly intolerant to delay, such as remote surgery for example.

6G, in addition to being seen as more than "5G on steroids" (Gatherer, 2018), relies heavily on the use of artificial intelligence (AI) (Saad et al., 2020). In fact, the synergy between AI and 6G is expected to open future horizons in an unprecedented way. Not only AI will be used extensively in the network to have an optimized 6G performance, but also 6G will provide the necessary infrastructure for the exploding use of AI in almost every sector: healthcare, transportations, industry, etc. New use cases, consisting of the combination of two or more of the 5G use cases, will emerge, such as mobile broadband reliable low latency communication (MBRLLC), where the service requirements are the union of those of the eMBB and URLLC use cases, and massive URLLC (mURLLC), where the service requirements are the union of those of the mMTC and URLLC use cases (Saad et al., 2020; RF Wireless World, 2020).

These 6G use cases, along with the advancements in AI, will unlock a variety of novel services, such as "extended reality" (XR), "connected robotics and autonomous systems" (CRAS), and "human-centric services" (HCS). These services would not have been possible without the "convergence of communications, computing, control, localization, and sensing (3CLS)" (Saad et al., 2020) enabled by the 6G-AI synergy.

It should be noted that such advanced services depend on the deployment of adequate infrastructure that can their support the needed requirements. Such infrastructure is available mostly in urban areas (Onireti et al., 2016), but significant connectivity gaps exist between these areas and the rural areas of developing countries (McKinsey, 2014). Global initiatives have emerged with the aim of bridging, or at least reducing this gap, such as "basic internet connectivity" or "global access to the internet for all (GAIA)" (Onireti et al., 2016), with the aim of providing "affordable broadband" in these areas (Mekuria and Mfupe, 2017). Hence, although the full-fledged 6G connectivity will not be achievable in these areas, several initiatives and techniques can be adopted to provide a certain degree of connectivity, thus leading to digital inclusion (Dixit, 2019). These techniques are surveyed in (Yaacoub and Alouini, 2020), and rely mostly on satellite and unmanned aerial vehicle (UAV) platforms (Giordani and Zorzi, 2019; Giordani and Zorzi, 2020). They are aligned with the United Nations (UN) sustainable development goals (SDGs) (Latva-aho and Leppänen, 2019). The human centric aspect of 6G was also stressed in (Dang et al., 2020).

In this paper, the new horizons unlocked by the synergy between 6G and AI are overviewed, and the risks faced are discussed. Several papers in the literature have provided surveys on the road to 6G and discussed the importance of AI in 6G, e.g., (Dang et al., 2020; Zhu et al., 2020; Jiang et al., 2021; Bin Ahammed and Patgiri, 2020). These papers provide valuable insights by focusing mostly on the technical requirements and characteristics that enable a successful 6G deployment. This article avoids repeating the same analysis by focusing more on the applications and real-life scenarios that can be enabled when 6G/AI are deployed and the technical characteristics described in the literature are met.

This paper is organized as follows. In Section 2, open horizons enabled by 6G and AI are discussed. They correspond to scenarios that, until recently, were treated as part of the realm of science fiction. However, with 6G and AI, they seem to be within reach in the next decade or so. Section 3 analyzes the security risks that affect and hinder the realization of the scenarios of Section 2. Finally, Section 4 concludes the paper.

## 2. HORIZONS

This section describes some important new horizons unlocked by the synergy between 6G and AI. The focus is on seven main topics, shown in Figure 1, that are attracting significant research interest, and are expected to do so in the near future.

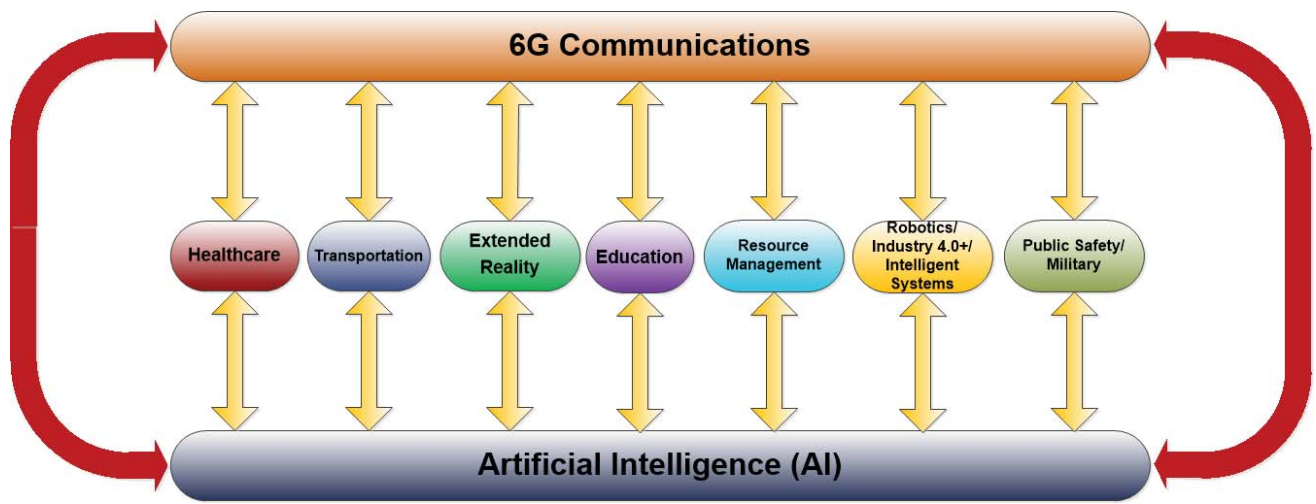

Figure 1. Main topics discussed in this paper, unlocked by the 6G-AI synergy.

### 2.1 Healthcare

Mobile health (mHealth) techniques are already widely deployed currently. Wearable sensors are used to collect data, and various edge/cloud techniques are used to process it, analyze it, store it, and exploit it. Hence, AI is widely used in healthcare and ambient assisted living to monitor patients, detect illnesses/anomalies, and notify medical response teams accordingly (Yaacoub et al, 2020b).
3

With 6G, this process is becoming even more pervasive. In conjunction with AI, continuous measurements of health parameters and vital signs could allow the generation of a "digital twin" of the monitored person. Advanced AI techniques implemented on the cloud could then perform heavy computations to identify any illnesses, or simply to provide recommendations on healthiest lifestyle to increase the longevity of that specific person (Arias Garcia and Roseman, 2021). A risky surgery can be attempted "virtually" on the digital twin as many times as needed until it succeeds, before performing it on the actual patient. The "digital twin" can be enabled by more extensive and accurate measurements, e.g., through the use of Internet of Nano-Things (IoNT) medical sensors/actuators inside the body (Sicari et al., 2019). Advances in this area could enable revolutions in precision medicine. For example, "bionauts" could travel inside the body and inject chemical treatment directly on cancerous cells, instead of having traditional chemotherapy target both healthy and cancerous cells (Diamandis, 2021).

IoNT mainly uses two communication types (Sicari et al., 2019):

- Nano scale communications using electromagnetic waves, mostly terahertz communications.
- Nano scale communications based on the exchange of molecules in the body (molecular communications).

Molecular communications are slower and subjected to various biological parameters that can affect them. They are less understood than electromagnetic communications, where channel models have been investigated for the propagation of these signals in the blood (Jornet and Akyildiz, 2011). Thus, these waves would be easier to deal with when designing security schemes for IoNT in human body.

The communication techniques are used between the various constituents of an IoNT, namely (Mezaal et al. 2018; Sicari et al., 2019):

- Nano-devices or nano-nodes: these are the simple low complexity devices responsible for sensing the environment and collecting relevant data.
- Nano-routers: these are more powerful devices that receive the transmissions of the nano-nodes and aggregate them. They can implement more advanced security mechanisms.
- Nano-controllers: These devices receive the data from the nano-routers and forward it to gateways.
- Gateways: They receive the data from controllers and can send it to cloud and/or edge servers for storage and processing

An example of the deployment and interaction between these various devices is shown in Fig. 2.

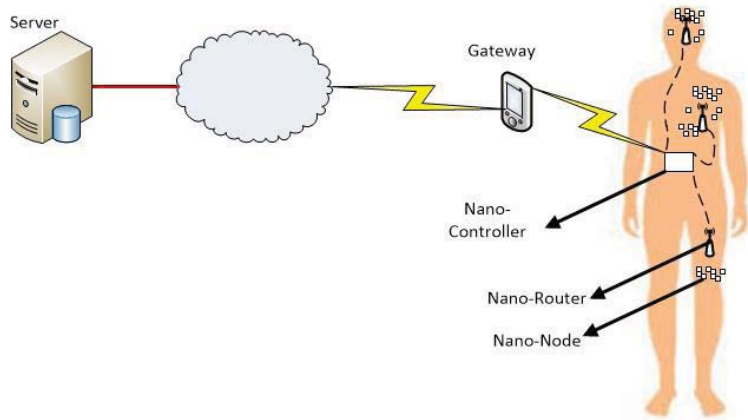

Figure 2. IoNT devices in a health scenario. Terahertz communication used between nano-nodes is one of the key technologies investigated in 6G.

It should be noted that communications between controllers and gateways can generally be secured using traditional security methods used for IoT, e.g., (Yaacoub et al, 2019; Yaacoub et al, 2020a; Yaacoub et al, 2020b).

The more challenging innovations need to be performed for securing the communications between the nodes, routers, and controllers. In practice, it would be very hard for an attacker to physically access the nodes or be too close to them. Moreover, their weak signals make them a difficult target for eavesdropping (Sicari et al., 2019). However, compromising the router or controller can cause serious threats. For example, controlling these routers and/or controllers could in turn lead to using them for controlling the nano-sensors/actuators, which not only could lead to network malfunction or communication disruption, but also could affect the life of the person/patient by altering the operation of these devices in the biological environment (Akyildiz et al., 2015; Atlam et al., 2018; Akhtar and Perwej, 2020).

On the cloud side, assuming zero trust, federated learning can be used to protect the patients' privacy, thus allowing the transmission of the weights of the machine learning system instead of transmitting the whole data. This assumes the presence of some intelligence at the edge (to perform training at the patient's side) and is not applicable in scenarios where a digital twin is to be created at the cloud (since actual data is not transmitted to the cloud). Moreover, measures need to be taken to identify and mitigate the effect of malicious nodes who send fake weights to the server in order to tamper with the server's result that are communicated to the legitimate nodes (Gouissem et al., 2022).

## 2.2 Transportation

The advances in intelligent transportation systems have been largely reported in the research literature of the last two decades. Connected cars, V2X connectivity, and self-driving cars have been studied extensively. The synergy between AI and 6G is expected to take this research into the next level, and allow large scale adoption of these technologies (Mchergui et al. 2021; Marr, 2021; Pao, 2021).

In fact, better AI leads to better and safer safe driving cars/vehicles. With 6G connectivity, these self-driving vehicles can communicate faster and respond with ultra-low latency to emergency events. Once legal and psychological concerns are overcome, this would allow a vast adoption of self-driving cars and autonomous vehicles.

Moreover, new business models could emerge from this process. If a person returns home from work and is spending the evening at home, then instead of having their self-driving car locked in the parking, it can be put to better use: It can be used as a rental, or as a taxi (sort of a driverless Uber), that can carry passengers while the owner is sitting in the comfort of his/her home. Hence, cars can make money for their owners when they are not using them (Muoio, 2016). Ordering and dispatching of cars would be done through 6G connectivity, whereas AI can be used to perform trajectory optimization, finding the nearest car that fits certain criteria requested by the passenger, alert the owner to perform preventive maintenance by monitoring the car parameters, etc.

In the same way, this can revolutionize the supply chain and delivery networks. Self-driving trucks could be sent on optimized route by a central AI. Their position could be monitored at every instant, and their merchandise could be emptied by robots waiting for them at destination warehouses. 6G connectivity allows tracking of every single goods item at every step of this process. A similar approach is also expected even for boats (autonomous shipping) (Marr, 2021; Liu et al, 2022).

Naturally, as mentioned previously, legal and psychological hurdles still need to be overcome. For example, some people might not feel safe when in a self-driving car, as they prefer to have control. Also, if a self-driving car makes an accident, who will be held legally liable: the car owner, the car manufacturer, or the car itself? Nevertheless, advances in AI and 6G connectivity are expected to minimize the risk of accidents, optimize traffic routes, and increase the revenues of car owners (while disrupting the taxi industry).

*2.3 XR*

With the capabilities of 6G and XR, the current virtual meetings in 5G can seem as a relic from the distant past. The real-time MBRLLC 6G capabilities will make the virtual environment more realistic. Holograms and haptics can allow a virtual handshake to feel as a real one. Advances in natural language processing (NLP) can allow for immediate translation, such that a virtual meeting with international participants, each speaking a different language, can appear to each participant as taking place completely in his/her own language. These advances in haptics, 6G and VR can similarly make romantic encounters in long-distance relationships feel like face-to-face dates (Tong and Zhu, 2021). Coupling this with AI and digital twin techniques, these encounters could even happen with holograms of deceased loved-ones, with their avatars summoned from the cloud, where their digital twin is stored (Banham, 2019). Moreover, digital twin in healthcare allows surgeons to practice a risky surgery multiple times on the patient's digital twin, before performing it on the actual patient. Fig. 3 shows a schematic of this scenario, which also includes vital signs monitoring of the surgeons, to make sure they can perform the surgery with sufficient level of focus while not being too stressed. Once ready, they can operate on the patient.

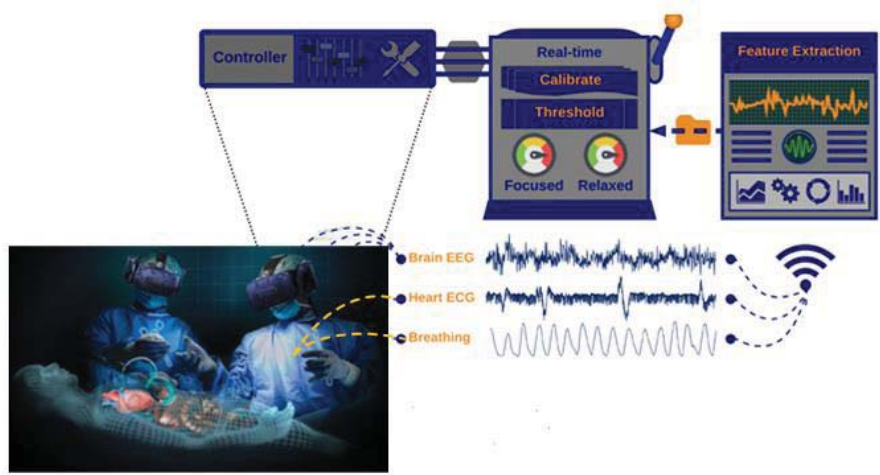

Figure 3. Example of surgeons practicing surgery on a digital twin using virtual reality.

The tourism industry can benefit tremendously from such advances, through the use of augmented reality. For example, a visit to a historical castle can be augmented by "meeting" avatars of the inhabitants who lived there centuries ago. Moreover, one can have virtual tours or virtual "trips" from the comfort of their home, where VR would allow them to be immersed in the destination of their choice before deciding to make the actual trip (Messier, 2016). Advances in haptics and e-touch/e-smell, coupled with 6G MBRLLC communications, can allow even having a taste of the local cuisine, or enjoying the aroma of local flowers, for example. Similarly, with 6G, haptics, and XR, people can attend an event or concert without physically being there. The entertainment industry could thus have different types of tickets for physical and virtual attendance.

XR is also expected to revolutionize the retail industry. Shopping can be done at home, where an immersion in a virtual shop allows a shopper to try dresses on her avatar, which has similar body measurements calculated from IoT wearable sensors. With 6G connectivity, a literal "blink" in the head mounted display or VR glasses allows the purchase to be made, the payment being charged to the shopper's credit card, and a drone can deliver the product the next day at the shopper's doorstep (Diamandis, 2019).

Thus, the progress in this area is expected to attract more people to spend more time immersed in virtual environments, which raises concerns about the blurring of boundaries between real and virtual worlds (Tegmark, 2017).

## 2.4 Education

The advances in technology will allow classrooms of the future to bypass the traditional education systems and traditional learning techniques. The adoption of modern digitally-enriched techniques will allow the students to learn more efficiently and be better equipped to face modern challenges and contribute into society and the economy (Deloitte, 2017). In addition to the now common use of tablets, laptops, interactive boards, the adoption of VR/AR applications will be essential in these enhanced education systems (Deloitte, 2017).

Beyond the use of VR/AR techniques as a learning tool, they can also be used to increase the students' attention and concentration levels. Indeed, electroencephalography (EEG) headsets have now evolved and are mostly using non-invasive electrodes. In addition, the brain signals they measure are transmitted wirelessly without the use of cumbersome wires. Furthermore, they can be used in conjunction with VR headsets in a combined and integrated fashion, e.g., (DSI VR 300, 2022). With edge processing, there is not even a need to transmit the EEG signals themselves. They can be processed locally and analyzed with adequate AI algorithms, and only the values of the metrics related to focus/attention/concentration could be sent to the teacher (Hashash et al., 2021).

With this technology, biofeedback, or more specifically NeuroFeedback (NFB) in the case of EEG, can be implemented. Using EEG NFB, brain waves are measured and a feedback signal (generally an audio-visual signal) is provided through the VR headset in order to train subjects to, for example, increase their concentration and focus levels (Marzbani et al, 2016). Using this technology, the learning process can be significantly enhanced. Less-performing students could be identified and would receive the needed assistance. Moreover, electrocardiography (ECG) signals could be used to detect the stress levels (Healey and Picard, 2005) of the students during exams or during the explanation of challenging material. ECG is the process of recording the electrical activity of the heart over a period of time using electrodes placed on the skin. Recent advances with the photoplethysmography (PPG) technique allow measuring the heart rate without the need to place electrodes on the skin. This technique allows the detection of the pulse wave travelling through the body by detecting the changes in the amount of blood in a specific body part by using light transmission through a tissue (e.g. a finger) and then capturing the light through a probe (Jensen and Hannmose, 2014). Simple techniques for measuring the heart rate can now be performed using a smart-watch (Cronovo, 2022) or a smart ring (Oura Ring, 2022). Hence, ECG signals can be monitored through simple wearable devices and the levels of stress can be reported back to the instructor. Figure 4 illustrates the scenario of immersive education with ECG measurement.

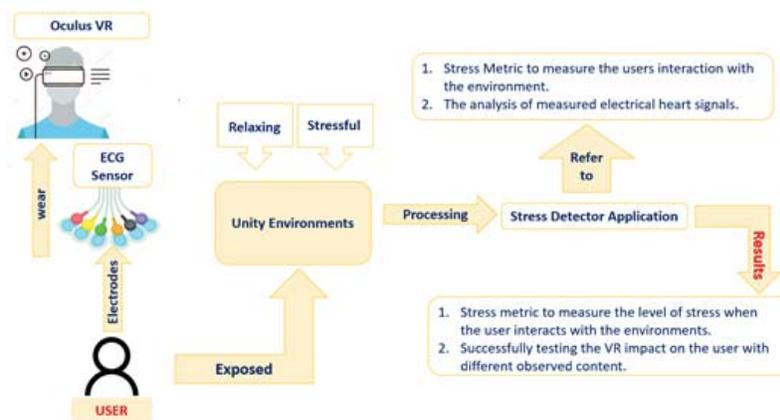

Figure 4. ECG measurement while the student is immersed in a VR educational environment.

Thus, 6G allows online education to be more pervasive, with support for VR/XR applications that can provide a virtual classroom environment where a student's can attend the class with avatars of their colleagues. In such an environment, AI, through the use of IoT for vital signs monitoring, can allow not only the detection of engagement levels of students, but also the adoption of necessary measures through biofeedback. In fact, depending on the EEG/ECG measurements, the educational content can be adapted to the needs of a particular student, or the teacher can interfere to take appropriate action and help increase the attention and/or reduce stress of certain students.

Although the current cost of certain equipment needed for the above scenario is high, e.g., (DSI VR 300, 2022), it is expected to decrease with technological process and market adoption. In addition, other less costly EEG devices using less electrodes, e.g. see (MUSE, 2022), can be considered as consumer electronics and are commercially available. They can be used by students at home with a mobile application and NFB provided through music, in order to improve their concentration before studying. These simplified EEG devices (MUSE, 2022) can also be used in conjunction with a separate VR headset.

The approach described above can be used for not only education, but also for employee training or continuous learning. In fact, one of the societal challenges due to exponential progress in technology is that innovation might destroy jobs in some sectors but create jobs in others. The challenge is in meeting the demand for the skill sets required in the new jobs. Thus employees should embark on a life-long learning style and educational systems should be reformed to be able to cope with technological advances. The problem here is that low-skilled workers might be strongly affected, whereas medium-skilled workers would be more able to enhance their skills and move to jobs requiring higher skills. The educational enhancements discussed previously should allow the educational system to be following closely the technologies advances.

## 2.5 Resource Management/Optimization

This section describes the use of AI and 6G to optimize resource management, where "resource" is used in a very general sense and could refer to energy, water, oil and gas, or the resources of the 6G network itself (Jamil et al., 2020).

In fact, not only 6G is serving AI to address the verticals discussed in this paper such as healthcare and transportation, but also AI is serving 6G to optimize the network performance. This can include radio resource management, network slicing, and optimized power consumption for green networking, among other issues (Du et al., 2020; Lee et al., 2021; Manogaran et al., 2021; Zeb et al., 2021). In fact, the literature is abundant with research works covering the implementation of the various types of machine learning (supervised, unsupervised, and reinforcement learning) to solve a multitude of 6G networking problems. Table I provides a brief non-comprehensive summary of recent publications using AI to optimize the performance of 6G networks.

Table I. Summary of AI Algorithms and their 6G Applications.

| Reference | Learning Technique | Algorithm | Target Solution in 6G |
|---|---|---|---|
| Ren et al., 2019<br>Jamshed et al., 2021 | Unsupervised Learning | K-Means Clustering | Resource allocation in uplink non-orthogonal multiple access (NOMA) with electromagnetic emission awareness |
| Chi et al., 2020 | Unsupervised Learning | K-means and clustering algorithm perception decision-CAPD | Reducing non linearity in 6G visible light communications (VLC) systems |
| Sritharan et al., 2020 | Supervised learning<br>Unsupervised learning<br>Reinforcement learning | Deep Neural Network<br><br>Deep Q-Learning | Resource management for data rate maximization |
| Salh et al., 2021<br>She et al., 2021 | Supervised learning<br>Unsupervised learning<br>Reinforcement learning | Deep Learning | URLLC Communications |
| Kim et al., 2021 | Supervised learning<br>Reinforcement learning | Deep Neural Network (DNN)<br>Convolutional Neural Network (CNN) | Medium Access Control (Resource allocation and random access) in 6G |
| Sami et al., 2021 | Reinforcement learning | Deep Reinforcement Learning | Resource provisioning for the Internet of everything (IoE) |
| Wang et al., 2021 | Reinforcement learning | Deep Reinforcement Learning | Handover in 6G VLC |
| Wu, 2021 | Reinforcement learning | Deep Q Network and Dueling Deep Q Network | Channel Monitoring |
| Shah et al., 2021 | Reinforcement learning | Hierarchical deep actor-critic networks | Network Control and Resource Allocation for 6G Space-Terrestrial Integrated Network |
| Zhao et al., 2021 | Reinforcement learning | Deep reinforcement learning scheme with sequential actor-critic model | Reducing delay and error rate in URLLC |

| Mei et al., 2021 | Reinforcement learning | Modified deep deterministic policy gradient (DDPG) and double deep-Q-network algorithm | Spectrum efficient network slicing in 6G |
|---|---|---|---|
| Cao et al., 2021 | Reinforcement learning | Federated deep reinforcement learning and deep Q networks | Access and handover control in open radio access network (O-RAN) |

Moreover, the mURLLC use case of 6G allows gathering huge amounts of data, while AI allows analyzing it and taking efficient action in real-time. For example, the smart power grid can become more intelligent by relying on AI to maximize the use of renewable energy sources in the energy mix while minimizing reliance on fossil fuel, to predict the load of the power grid and perform appropriate balancing, to take self-healing and self-optimization measures without human intervention, and to perform dynamic pricing based on the energy mix, consumer demand, and other economic considerations (Esenogho et al., 2022). The transmission of suitable commands across the network will critically depend on the performance of 6G mURLLC. The same principle applies to other utilities, where monitoring of water distribution networks can be performed similarly, with AI used to detect, or better yet, predict leakages through continuous monitoring of the state of the water pipes. The use of smart water or electricity meters, along with AI algorithms, can allow better understanding of user behavior, better prediction of network load, and better use of current network resources with optimized planning of network future expansions.

Similar approaches, if judiciously applied (IoT data collection, 6G transmission of massive amounts of sensor reading, AI analysis/prediction/forecasting and suggestion of solutions), could revolutionize many fields, e.g., farming (Usman et al., 2022), or lead to reducing pollution and slowing the global warming process, thus also safeguarding natural resources of the planet, when applied for environment monitoring.

The concept of edge computing is essential to the success of these techniques. With edge computing, part of the intelligence is moved from the cloud to be placed at the access network closer to the user. This allows faster computation, reduces the load on the network bandwidth, and increases energy efficiency (Loven et al., 2019). AI needs to cope with this architecture, moving from cloud-based AI in 4G, to enhanced AI services with some edge applications in 5G, to a fully edge-native AI in 6G, where self-learning AI for edge-computing is seamlessly integrated with the network (Xiao et al., 2021).

*2.6 Robotics/Industry 4.0+/Intelligent Systems*

The permanent connection of robots to the cloud through highly reliable, continuously available 6G networks allows them to share their learning experiences. Hence, when a robot self-learns something, the knowledge is instantaneously shared with all connected robots. Thus, 6G allows these AI systems to evolve exponentially (Tegmark, 2017; Dupont et al, 2021; Murphy, 2021).

With XR, IoT, and 6G, intelligent systems can be run even more intelligently. For example, an IT technician can wear a VR headset, and see a virtual rendering of the network. He can virtually navigate in an immersive environment, where buildings represent servers and network equipment; roads represent interconnections between them, and where the configuration parameters or performance indicators of each device can be inspected in the virtual world. This allows troubleshooting to be faster, easier, and more efficient (Ryan, 2017).

Similarly, the electric grid or an oil and gas pipeline network can be rendered in a similar fashion. The real-time parameters can be obtain from IoT sensors and reported in the virtual model for troubleshooting. Another approach is to combine real-time IoT readings with real images without the monitoring/troubleshooting team having to leave headquarters. For example, a drone can fly over an oil pipeline sending a video in real-time. The monitoring team can watch a 3D version of this video in an immersive environment, where augmented reality is used to overlay the readings of IoT sensors over the corresponding part of the pipeline. Thus, an engineer can see, as he "flies" over each pipe section, the pressure parameters, temperature, etc., corresponding to this specific section. AI can be used to provide preventive maintenance by analyzing the collected and stored IoT data, thus identifying potential risks and requesting the drone flights to be directed towards corresponding zones.

## 2.7 Public Safety/Military

The mastery of AI and 6G is expected to lead to remarkable advances in military prowess. The use of drones and UAVs can fundamentally change law enforcement operations and battlefield tactics. Figure 5 shows an example with UAV swarms, forming flying ad hoc networks (FANETs), can support the operation of ground troops (which can be equipped with IoT sensors and perform device to device (D2D) communications). The FANETs could also provide 6G connectivity with satellites.

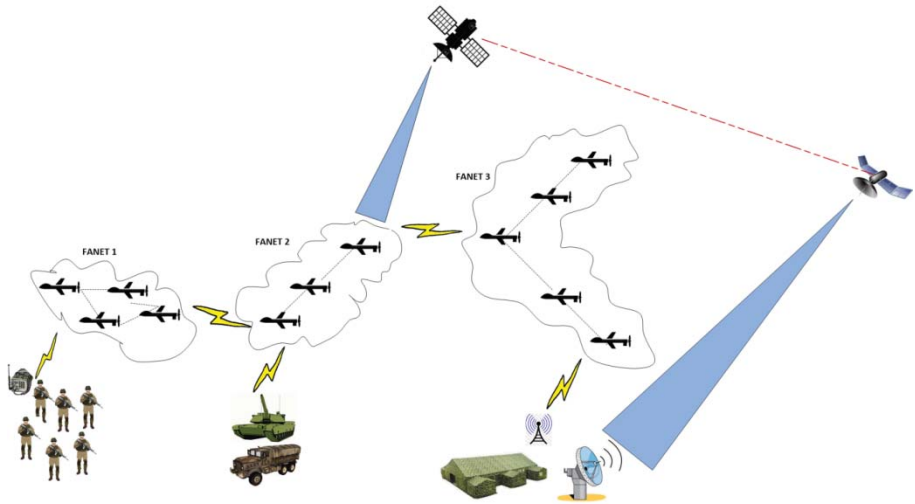

Figure 5. Military scenario with ground, UAV, and satellite communications.

Advances in robotics/AI can lead to robot soldiers, thus limiting human casualties while inflicting more damage on the enemy. Militaries using these advanced technologies will have a deterring force that can tip the balance of power in their favor. On the optimistic side, when used wisely by responsible actors, the mere display of this power can stop unnecessary wars (Reese, 2018). On the pessimistic side, the fall of this technology in the wrong hands can lead to dramatic consequences. For example, a nano-drone (in the form of a robotic bee or fly) can be used to track, find, and assassinate subjects (Tegmark, 2017). AI-optimized weapons in the hand of international terrorist organizations can lead to devastating damage. Several prominent AI scientists and researchers have signed a petition to ask governments to refrain from supporting AI research in the military domain (Tegmark, 2017; Reese, 2018). Even if their calls are heard, this does not guarantee their success. In fact, even if governments comply, progress in communications and AI in the other "civilian" areas discussed in the above subsections areas can be used to advance research in the military domain by malicious/terrorist groups, who of course do not feel obliged to respect the scientists' petition (Reese, 2018).

## 3. RISKS

The opportunities and future horizons discussed in the previous section provide a large target base for a wide range of security attacks and vulnerabilities. In fact, although AI techniques can be used to detect and mitigate security vulnerabilities and intrusions, advances in AI can also be used for perfecting the attacks targeting the systems described above. Thus, it can be a double-edged sword used for both attack and defense (Siriwardhana et al., 2021; Buchholz et al., 2022).

### 3.1 Malware and Attacks

Traditional malware and attacks are expected to still exist after the proliferation of 6G and AI. Viruses, worms, Trojan horses, denial of service (DoS) attacks, distributed DoS (DDoS) attacks, etc., will continue to exist. The progress in cybersecurity research and innovative countermeasures will limit their occurrence probability. However, in case of success, their impact and the damage they will cause will be devastating, and orders of magnitude larger than their current level of damage (Siriwardhana et al., 2021).

For example, due to their limited power and computational capabilities, IoNT devices face severe threats such as lack of encryption, vulnerability to a denial of service attacks, malware, etc. In fact, bio-cyber attacks over IoNT networks can be used to steal personal health-related information, and viruses can be used to hack deployed IoNT to cause serious health risks. Controlling the routers and/or controllers in a medical IoNT network could in turn lead to using them for controlling the nano-sensors/actuators, which not only could lead to network malfunction or communication disruption, but also could affect the life of the person/patient by altering the operation of these devices in the biological environment (Akyildiz et al., 2015; Atlam et al., 2018; Akhtar and Perwej, 2020). This is a kind of bio-cyber terrorism that not only steals personal health information but could also lead to murder.

Therefore, efficient and innovative methods are needed to maintain the confidentiality, integrity, and availability of the data. These could include, for example:

- Efficient key management and distribution techniques between gateways, nano-controllers, and nano-routers to perform symmetric encryption quickly and efficiently, with reasonable computational overhead. A potential solution could be the use of quantum key distribution between the server and gateway, which would then handle the transmission of the generated keys of Terahertz wireless channels to the controllers and routers.
- Physical layer security techniques to add an extra layer of security. This includes investigating physical layer techniques as an add-on to encryption at the nano-routers and controllers, and as a substitute for encryption at the nano-nodes - router links. In fact, for this latter scenario, the limited capabilities of the node make this approach desirable. However, its implementation within a biological propagation environment would be challenging. A more challenging and novel scenario would be to use molecular communications as an additional protection layer, in conjunction with security measures at the terahertz wireless level.
- AI-based techniques for intrusion detection and mitigation of denial-of-service attacks targeting the gateway, controllers, or routers. As mentioned previously, any breach at this level can lead the malicious attacker to control the nano-nodes and cause severe damage.

As another example, hacking the control system of the power grid could lead to system instability (Tariq et al., 2020), and possibly plunge entire nations in darkness. On a smaller and less dramatic scale, hacking smart meters could lead to scenarios where a person's power consumption is billed to their neighbor.

In the self-driving cars scenario, the cars would be expected to be connected to cloud servers (or some remote servers of their manufacturer) for software upgrade, preventive maintenance, etc. A breach in these servers could allow the hacker to remotely control millions of cars at the other end of the world. Imagine a scenario where empty cars are swarming in the streets, closing major road intersections, blocking movement, while being controlled by an unknown malicious user.

The previous examples showcase the importance of embedding security measures in the design of AI and 6G systems, and allowing the option to return to more "primitive" techniques in case of disaster. This could involve, for example, allowing drivers to return to manual mode by disconnecting the cars (e.g., pressing a special button), or disconnecting the power grid from the telecom network and using manual techniques to route the power through the distribution networks. This requires to still have employees trained in the "old ways" in addition to the more tech savvy ones running the "intelligent" grid.

### 3.2 Eavesdropping/Jamming

These physical layer techniques are expected to become more efficient in the 6G/AI era (Nguyen et al., 2020). In fact, the strong reliance on small IoT devices presents an obvious weakness (Dastjerdi and Buyya, 2016; Al-Garadi et al., 2020): Due to their limited capabilities, these devices cannot resist strong jamming attacks. These attacks, could for example, disable the operation of actuators in a medical IoNT network, thus disrupting medicine delivery by bionauts. Moreover, the reliance of IoT sensors to relay all kinds of measurements allows an eavesdropper to gain valuable information. It should be noted though that eavesdropping in certain scenarios, like the IoNT nano-sensors, is difficult since the devices power capabilities are too weak and an eavesdropper would have to be physically present near the patient in order to capture meaningful data.

In general, physical layer techniques could be used to mitigate these attacks, e.g., noise injection to combat eavesdropping or beamforming (whenever supported by the devices) to mitigate jamming, and generation of encryption keys by relying on physical channel parameters, such as channel state information (Nguyen et al., 2020). Moreover, hybrid techniques, involving the use of encryption and more traditional security methods in conjunction with physical layer techniques, can be used to mitigate such attacks. Quantum key distribution, although cannot be supported by small IoT devices, could be used by routers/controllers to provide strong cryptographic keys for IoT devices, as discussed in the next section (Al-Mohammed at al., 2021).

### 3.3 Quantum Computing

The progress in quantum computing, although generally beneficial to AI techniques due to speeding up computations, is also beneficial to malicious attackers trying to detect encryption keys. In fact, it becomes important to design encryption techniques that can hold long enough in the presence of attackers equipped with quantum capabilities. The massive deployment of IoT devices does not help in this direction (Lohachab et al., 2020; Fernández-Caramés, 2020).

A possible solution is to adopt quantum key distribution (QKD) by the IoT controllers. Being more powerful devices, they can be connected to servers over a fiber optic connection, in addition to using a traditional radio frequency (RF) wireless connection. They can benefit from the optical connection to use photons for generating quantum keys. Then, they can distribute these keys to the IoT devices over the local wireless connection according to some secure distribution architecture or protocol. This process can be repeated periodically to void using the keys for too long and risking to compromise the communications (Al-Mohammed at al., 2021).

AI techniques can also be used to detect the presence of quantum attackers performing man-in-the-middle attacks to detect the quantum keys being exchanged as part of QKD (Al-Mohammed at al., 2021).

*3.4 Zero Trust*

In the open radio access network (ORAN) of 5G+/6G, a zero-trust architecture is adopted. This requires security measures designed to operate in a way as if the attacker is already inside the organization or network (Nguyen et al., 2021). This is important in today's and future networks with the billions of connected devices, cloud access, and critical data being circulated, processed, and analyzed. Frequent authentication should be used for users, applications, and infrastructure devices. In addition to confidentiality, integrity, and authentication, this raises the issue of trustworthiness, which must be tracked in real-time according to certain metrics for each device in the network (Nguyen et al., 2021; Ziegler et al., 2021; Tong and Zhu, 2021). Thus, every operation should be properly saved in the logs for further inspection. This could be done continuously (by appropriate AI techniques) to detect any threats, or after the occurrence of any incident through digital forensics techniques.

*3.5 Blockchain*

Blockchain has gained significant interest in privacy and security research, in addition to research on its role in virtual currencies. Consequently, it was proposed for providing privacy and security in 6G networks (Hewa et al., 2020; Sun et al., 2021; Velliangiri et al., 2021). Blockchain is spanning the whole range of 6G-related security research, from securing edge data (Sun et al., 2021), to securing the 6G access network itself (Velliangiri et al., 2021), to complete vertical sectors such as healthcare, Industry 4.0+, and environment monitoring (Hewa et al., 2020).

However, although 6G could provide the connectivity required for blockchain traffic, several challenges seem to be either ignored, or at least underestimated by most of the literature. For example, will the scalability of blockchain hold with millions of transactions? Since the blocks are hashed and interconnected, how long would it take for a specific transaction be retrieved and decrypted if one has to go all the way back through the chain in order to reach it? Thus, having shorter chains might solve this problem, but will reduce the security of the blockchain approach. Moreover, the energy required for transaction validation, similarly to bitcoin mining, is significant. The consumption and cost will increase with the number of nodes added to the blockchain (Ali, 2021). In fact, the energy needs of blockchain, coupled with the sensitivity of blockchain to energy prices and availability, e.g., during the crisis in Kazakhstan at the beginning of 2022 (Nahar, 2022), should raise some flags about the large-scale adoption of blockchain in 6G.

## 4. CONCLUSIONS AND DISCUSSION

In this paper, the synergy between 6G and AI was investigated. Seven areas of high research interest were overviewed, and it was argued that, although until recently some of their aspects were closer to the realm of science fiction, the 6G-AI synergy will make them within reach. However, excessive reliance on connectivity and AI comes with an increased threat level and serious risks that cannot be overlooked. Consequently, the paper also discussed the most important security challenges involved and suggested potential mitigation techniques.

Future research directions are expected to continue in using AI to optimize the performance of 6G networks, and in the use of 6G-AI synergy to reach further achievements in the various vertical areas discussed in the paper (healthcare, transportation, education, etc.). However, an important barrier to achieving the future horizons discussed in this paper is not the technical limitations, which are being, and will generally be, overcome. It is rather reaching a wide acceptability of the role of AI in our lives. For example, it will be hard for a patient, or a physician, to accept a decision made by AI in a healthcare monitoring scenario, based only on the data without a convincing reason or demonstration of symptoms. Similarly, psychological barriers (more than technical ones) need to be overcome before seeing wide adoption of driverless taxis or airplanes. Therefore, more efforts need to be spent, in addition to the existing ones, to perform more research on areas like "explainable AI", "ethical AI", "justifiable AI", in addition to legal aspects of AI, so that a certain maturity level is reached in these areas before, or at least in parallel to, future wide deployment of 6G.

## ACKNOWLEDGMENTS

This publication was jointly supported by Qatar University and IS-Wireless - IRCC Grant no. IRCC-2021-003. The findings achieved herein are solely the responsibility of the author.

# REFERENCES

Akhtar, N. and Perwej, Y. (2020). The internet of Nano Things (IoNT) Existing State and Future Prospects. *GSC Advanced Research and Reviews*, 05(02), 131–150.

Akyildiz, I. F., Pierobon, M., Balasubramaniam, S., and Koucheryavy, Y. (2015). The Internet of Bio-Nanothings. *IEEE Communications Magazine — Communications Standards Supplement*, pp. 32-40, March 2015.

Al-Garadi, M.A., Mohamed, A., Al-Ali, A., Du, X., and Guizani, M. (2020). A Survey of Machine and Deep Learning Methods for Internet of Things (IoT) Security. *IEEE Communications Surveys and Tutorials*, 22(3), 1646–1685.

Al-Mohammed, H., Al-Ali, A., Yaacoub, E., Qidwai, U., Abualsaud, K., Rzewuski, S., and Flizikowski, A. (2021). Machine Learning Techniques for Detecting Attackers during Quantum Key Distribution in IoT Networks with Application to Railway Scenarios. *IEEE Access*, 9, 136994 - 137004, doi: 10.1109/ACCESS.2021.3117405.

Ali, F. (2021). "The Top 5 Problems With Blockchain Technology", MakeUseOf.com, July 16, 2021 url: https://www.makeuseof.com/problems-with-blockchain-technology/ [Accessed: February 21, 2022].

Arias Garcia, D., and Roseman, J. (2021). "How Digital Twin Technology is Disrupting Healthcare," Plug and Play article, url: https://www.plugandplaytechcenter.com/resources/how-digital-twins-technology-disrupting-healthcare/   [accessed February 1, 2022]

Atlam, H. F., Walters, R. J., and Wills, G. B. (2018), Internet of Nano Things: Security Issues and Applications. Proceedings of the 2018 International Conference on Cloud and Big Data Computing. ACM. 7 pp. (doi:10.1145/3264560.3264570).

Banham, R. (2019). "The Departed: Communicating With Lost Loved Ones Through AI and VR", Dell Technologies, url: https://www.delltechnologies.com/en-us/perspectives/the-departed-communicating-with-lost-loved-ones-through-ai-and-vr/ [Accessed: February 1, 2022]

Bin Ahammed, T., and Patgiri, R. (2020). "6G and AI: The emergence of future forefront technology," in Proc. advanced communication technologies and signal processing conference, ACTS 2020, Silchar, India, Dec. 4–6, 2020, 1–6.

Buchholz, S., Bechtel, M., and Briggs, B. (2022). "Deloitte Insights - Tech Trends 2022", url: https://www2.deloitte.com/us/en/insights/focus/tech-trends.html [Accessed: February 1, 2022].
20